\documentclass[aps,prl,showpacs,showkeys,twocolumn,amssymb,groupedaddress]{revtex4}

\usepackage{graphicx}
\usepackage{dcolumn}
\usepackage{bm}

\begin{document}

\title{Phenomenology of the muon-induced neutron yield}

\author{A. S. Malgin}
 \email{malgin@lngs.infn.it}

\affiliation{Institute for Nuclear Research of the Russian Academy of
Sciences \\
60-letiya Oktyabrya prospekt 7a, 117312 Moscow, Russia}


\begin{abstract}
The cosmogenic neutron yield $Y_n$ characterizes the matter ability to produce neutrons under the effect of cosmic ray muons with spectrum and average energy 
corresponding to an observation depth. The yield is the basic characteristic of cosmogenic neutrons. The neutron production rate and neutron flux both are  
derivatives of the yield.
The constancy of the exponents $\alpha$ and $\beta$ in the known dependences of the yield on energy $Y_n {\propto} E_{\mu}^\alpha$ and the 
atomic weight $Y_n {\propto} A^\beta$ allows to combine these dependences in a single formula and to connect the yield with muon energy loss 
in the matter. As a result, the phenomenological formulas for the yield of muon-induced charged pions and neutrons can be obtained. 
These expressions both are associated with nuclear loss of the ultrarelativistic muons, which provides the main contribution to the total 
neutron yield. The total yield can be described by a  universal formula, which is the best fit of experimental data.
\end{abstract}



\maketitle

\section {I. Introduction}
Cosmogenic neutrons are of interest as a source of background in the underground low-background experiments. Neutrons 
generated in the matter by cosmic-ray muons are considered cosmogenic. Neutrons generated by astrophysical, atmospheric and solar neutrinos are also cosmogenic. 
The term "cosmogenic" has been associated only with neutrons from muons by virtue of their dominant role in the flux of neutrons generated at depths 
of up to 10 km w.e. underground by particles coming from space.

Cosmogenic neutrons (cg-neutrons) can be characterized by the following values: a neutron yield $Y_n$ ($n/{\mu}$/(g/cm$^2$)), a production 
rate $R_n(H)$ = $I_\mu (H)Y({\overline E}_\mu)$ $(n$\ g$^{-1}$s$^{-1})$, and a neutron flux ${\Phi}_n(H)$ = $R_n(H) l_n{\rho}$ = 
$I_{\mu}(H) Y({\overline E}_\mu) l_n \rho$ ($n$cm$^{-2}$s$^{-1}$). In these expressions, ${\overline E}_{\mu}$ is the mean muon energy at a depth H; 
$I_\mu (H)$ is the muon global intensity; $l_n\rho$ (g/cm$^2)$ is an attenuation length for the isotropic neutron flux. 
The indicated characteristics allow to estimate the background effects caused by muon-induced neutrons in rock and set-up materials 
using Monte Carlo simulations which take into account configuration and dimensions of the target. 

As follows from the above expressions, the main characteristic is the neutron yield $Y_n$. The production rate $R_n$ and the flux $\Phi_n$ 
are the derivatives from the yield $Y_n$. The formula for the 
neutron yield in the material $A$ at muon energy $E_{\mu}$ is:
\begin{equation}
Y_n(A,E_\mu) = \frac{N_0}{A}{\langle}{\sigma}_{{\mu}A}{\nu}_n{\rangle}, \hbox{($n/{\mu}$/(g/cm$^2$))}, 
\label{eq1}
\end{equation}
where $N_0$ is Avogadro's number, ${\langle}{\sigma}_{{\mu}A}{\nu}_n{\rangle}$ is a mean value of the product of a ${\mu}A$-interaction cross-section 
and a neutron multiplicity ${\nu}_n$.
The multiplicity ${\nu}_n$ includes all the neutrons (including multiplication neutrons) which arise mainly in hadron and electromagnetic showers 
produced via ${\mu}A$-interactions and developed entirely in the matter.  So, the product ${\langle}{\sigma}_{{\mu}A}{\nu}_n{\rangle}$ is a neutron 
production function. The cg-neutron energy spectrum which corresponds to the yield (1) will not be considered here.

The yield $Y_n$ in a line with other physical properties of the matter presents an ability of the matter to produce neutrons under the effect of muons.
In \cite{Aga13_rus}, \cite{Aga13} a universal formula (UF) was obtained for the muon-induced neutron yield: 
$Y_n^{\mathrm{UF}} = b^{ \mathrm{UF}} E_{\mu}^{\alpha}A^{\beta}$, $b^{\mathrm{UF}} = 4.4{\times}10^{-7}$ cm$^2$/g, $\alpha$ = 0.78, $\beta$ = 0.95. 
The formula is valid in the energy range from ${\sim}$ 40 GeV up to the maximum mean muon energy underground ${\sim}$ 400 GeV.
The lower limit of the range corresponds to a depth of about 100 m w.e. This UF is the best approximation of the set of available experimental data (Table. 1). 
The UF was obtained under the assumption that the dependence of the yield on $E_{\mu}$ and $A$ can be expressed as $E_{\mu}^{\alpha}$ and 
$A^\beta$, where $\alpha$  and $\beta$ are constant. The coefficient $b^{\mathrm{UF}}= 4.4{\times}10^{-7}$ cm$^2$/g is close to the relative 
muon nuclear energy loss $b_n = 4.0{\times}10^{-7}$ cm$^2$/g. 
The UF effectiveness is shown in Fig.1. The set of points in the coordinates $Y_n^{\mathrm{UF}} - ^{ex}Y_n$ 
is aligned at angle ${\alpha} = 45^{\circ}$ to the x-axis ($^{ex}Y_n$ are experimental data from Table 1).
Obviously, if ${\alpha} = 45^{\circ}$ then $Y_n^{\mathrm{UF}} $ = $^{ex}Y_n$. Thus, the UF expression relates the yield to muon energy losses 
and nuclear properties of the matter.

The yield $Y_n$ at depths $H {>} 100$ m w.e. (${\overline E}_{\mu}{>}$ 40 GeV) is a sum of the components $Y_n^h$, $Y_n^{\mathrm{em}}$ and $Y_n^v$. 
The components $Y_n^h$ and $Y_n^{\mathrm{em}}$ correspond to the neutron production in hadron ($h$) and electromagnetic (em) showers. 
Components $Y_n^v$  and $Y_n^{\mathrm{em}}$  present mainly photoneutrons which are produced in giant dipole resonance (GDR) by virtual 
photons ($Y_n^v$ ) or real photons of em-showers ($Y_n^{\mathrm{em}}$ ). The contribution of neutron production via ${\mu}^-A$-captures 
at depths greater than 100 m w.e. is negligible.

The ratio of neutron production channels has being considered by many authors \cite{Rya65}, \cite{Gor72}. It was shown that at energies 
40 $\le {\overline E}_{\mu} \le$ 400 GeV 
the yield components for all $A$'s are connected by the inequalities:
\begin{equation}
Y_n^{\mathrm{em}} {\gg} Y_n^v, Y_n^h {>} Y_n^{\mathrm{em}} + Y_n^v. 
\label{eq2}
\end{equation}

\section {II. Neutron production in h-showers}

As follows from the inequalities (\ref{eq2}), neutrons from $h$-showers dominate in the total neutron yield. In the $h$-shower neutrons are produced 
mainly in deep-inelastic ${\pi}A$-interactions of charged shower pions ${\pi}_s^{\pm}$, as well as ${\pi}^-A$-captures. The $h$-shower structure also 
contains ${\pi}^0$ initiating the development of em-subshower. The number of neutrons in the em-subshower is small compared with the hadron component 
of $h$-shower. Therefore, one can neglect neutron production in the em-subshowers.

The concept of neutron production in $h$-shower is based on an idea of intranuclear nucleon cascade (INC). Neutrons in $h$-shower are divided by origin 
into those "cascade" ({\em cas}) and "evaporative" ({\em ev}). {\em Cas-}neutrons are produced in the fast phase of ${\pi}A$-interaction as a result 
of the development of INC initiated by nucleon recoil from deep-inelastic ${\pi}N$-collision within a nucleus.
{\em Ev-}neutrons appear in the subsequent ${\pi}A$-scattering phase. They are emitted by the excited residual nucleus $A_r^*=A - N_{cas}$, here $N_{cas}$ 
is the number of cascade nucleons coming out of nucleus $A$, $1{\le}N_{cas} {<}A$. In the fixed energy $h$-shower the number ${\overline N_{cas}}$  
in ${\pi}A$-interaction and the average number of ${\pi}A$-interactions depend weakly on $A$ \cite{Pau62}, \cite{Bar72}.
The average number of {\em ev}-neutrons in $\pi A$-collision ${\overline n}_{ev}$ depends on a set of residual nuclei $A_r$, which is characterized 
by an average value ${\overline A}_r$. Thus, the number of neutrons in 
$h$-shower ${\nu}_{\pi}^{\pm}n_n$=${\nu}_{\pi}^{\pm}({\overline n}_{cas}+ a{\overline n}_{cas}+ {\overline n}_{ev})$  is related to the nucleus $A$; besides, 
the average number of {\em cas}-neutrons in $\pi A$-collision ${\overline n}_{cas}$ is associated with a mother nucleus $A$ but value of 
${\overline n}_{ev}$ is associated with a nucleus ${\overline A}_r$. 

Multiplicity ${\nu}_{\pi}^{\pm}$ is the average number of  ${\pi}^{\pm}A$-interactions in a shower, which is equal to the number of charged pions 
in the shower and weakly depends on $A$.
The addend $a{\overline n}_{cas}$ ($a {\ll}1$) takes into account multiplication of cascade neutrons in their $nA$-collisions. For any $A$ 
the value of ${\overline n}_{ev}$ is approximately 2 times the value of $(1+a){\overline n}_{cas}$. \\

\begin{widetext}
\begin{center}
\begin{table}[t]
\centering
\caption{\label{tab:data}Measured neutron yield}
\begin{ruledtabular}
\begin{tabular}{@{}l*{15}{l}}

 &  &  & \multicolumn{4}{c}{\textbf{$Y_n{\times}10^{-4}, n/{\mu}/(g/cm^2)$}} \\
\textbf{Experiment, Ref.} & \textbf{${\overline E}_{\mu}, GeV$} & \textbf{H, m.w.e.} & \textbf{$Y_{LS}$} & \textbf{$Y_{Fe}$} & \textbf{$Y_{Cd}$ } &  \textbf{$Y_{Pb}$ } \\
\hline
\cite{Ann54}          & $10.0{\pm}6.3$ &      20            &     -                & $0.98{\pm}0.01$  &    -      & $2.43{\pm}0.13$    \\
\cite{Ber70}          & $10.0{\pm}6.3$ &      60            &     -                &     -            &    -      & $4.8{\pm}0.6$      \\
\cite{Gor71}          & $11.0{\pm}6.6$ &      40            &     -                & $1.32{\pm}0.30$  &    -      & $4.03{\pm}0.36$    \\
\cite{Her95}          & $13.0{\pm}7.2$ &      20            & $0.20{\pm}0.07$      &     -            &    -      &     -              \\
\cite{Boe00}          & $16.5{\pm}8.1$ &      32            & $0.36{\pm}0.03$      &     -            &    -      &     -              \\
ASD, \cite{Bez73}     & $16.7{\pm}8.2$ &      25            & $0.47{\pm}0.05$      &     -            &    -      &     -              \\
\cite{Gor71}          & $17.8{\pm}8.4$ &      80            &     -                & $1.69{\pm}0.30$  & $3.3{\pm}0.4$ & $5.66{\pm}0.36$ \\ 
\cite{Ber70}          & $20{\pm}9$     &     110            &     -                &     -            &    -      & $6.8{\pm}0.9$      \\
\cite{Gor68}          & $40{\pm}12.6$  &     150            &     -                & $3.31{\pm}0.96$  & $10.3{\pm}4.3$& $11.56{\pm}1.1$ \\
ASD, \cite{Bez73}     & $86{\pm}18$    &     316            & $1.21{\pm}0.12$      &     -            &    -      &     -              \\
\cite{Bly15}          & $89.8{\pm}2.9$ &     610            & $1.19{\pm}0.21$      &     -            &    -      &     -              \\
\cite{Gor70}          & $110{\pm}21$   &     800            &     -                &     -            &    -      & $17.5{\pm}3.0$     \\
ASD, \cite{Rya86}     & $125{\pm}22$   &     570            & $2.04{\pm}0.24$      &     -            &    -      &     -              \\
\cite{Abe10}          & $260{\pm}8$    &    2700            & $2.8{\pm}0.3$        &     -            &    -      &     -              \\
ZEPLIN-III, \cite{Rei13} & $260{\pm}32$   &   2850          &     -                &     -            &    -      & $58{\pm}2$          \\
LSM, \cite{Klu15}       & $267^{+8}_{-11}$&   4850          &     -                &     -            &    -      & $27.5^{+10}_{-7}$  \\
\cite{Ber73}            & $280{\pm}33$   &    4300          &     -                &     -            &    -      & $116{\pm}44$       \\
LVD, \cite{Aga13}       & $280{\pm}18$   &    3100          & $3.3{\pm}0.5$        &  $16.4{\pm}2.3$  &    -      &       -      \\
LVD, \cite{Aga15}       & $280{\pm}18$   &    3100          & $3.6{\pm}0.3$        &  $14.3{\pm}1.3$  &    -      &       -      \\
Borexino, \cite{Bel13}  & $283{\pm}19$   &    3800          & $3.10{\pm}0.11$      &     -            &    -      &       -      \\
LSD, \cite{Aga13}       & $385{\pm}39$   &    5200          & $4.1{\pm}0.6$        & $20.3{\pm}2.6$   &    -      &       -            \\
\end{tabular}
\end{ruledtabular}

\end{table}
\end{center}
\end{widetext}

\begin{figure}[!t]
\centering
\includegraphics[width=2.8in]{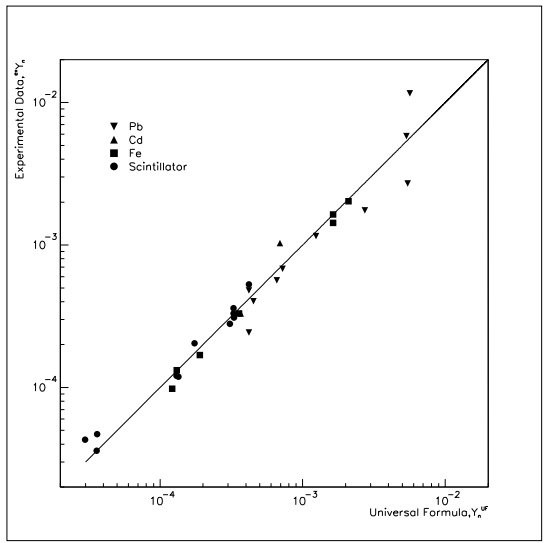}
\caption{Correspondence of the measured neutron yield $^{ex}Y_n$and the values calculated using UF.}
\label{1fig}
\end{figure}

\section {III. The neutron yield of charged pions and neutrons in h-showers} 

To describe the experimental data and to present the results of calculations of the yield $Y_n$ the following power-law dependences are used: 
\begin{eqnarray}
\label{eq3}
Y_n=c_AE_{\mu}^{\alpha} ({\mathrm{for~fixed~}} A),\\
\label{eq4}
Y_n=c_EA^{\beta} ({\mathrm{for~fixed~}} E_\mu),
\end{eqnarray}
where $\alpha$, $\beta$ are constant. The values of coefficients $c_A, c_E$ and exponents $\alpha$, $\beta$ are defined based on the best 
agreement  of the results of measurements or calculations with dependences (\ref{eq3},\ref{eq4}). They are adjustable parameters and have no physical meaning.
Simple dependencies (\ref{eq3},\ref{eq4}) at correct values $c_A, c_E, \alpha, \beta$ reflect well the tendency of the neutron yield change in a 
relatively small range of the mean muon energy underground from 40 to 400 GeV. Due to the constancy of exponents $\alpha$, $\beta$ and independency of $E_\mu$ and $A$
from each other we can factorize the $Y_n$ expression: 
\begin{equation}
Y_n=cE_{\mu}^{\alpha}A^{\beta}.
\label{eq5}
\end{equation}
In this case $c_a=cA^{\beta},c_E=cE_{\mu}^{\alpha}$, where $c$ is constant.

We can also use the power-law dependences (3,4,5) for the yield component $Y_n^h$.
Such a possibility is based on a broad experimental and theoretical material obtained in the early studies of multiple processes in hadron-nucleus 
collisions \cite{Bar72}, \cite{Mey63}, \cite{Koh67}, \cite{Bar52}. 

According to (\ref{eq1}), the yield $Y_n^h$ is given by
\begin{equation}
Y_n^h=\frac{N_0}{A}{\langle}{\sigma}_{{\mu}A}^h{\nu}_n^h{\rangle},
\label{eq6}
\end{equation}
here ${\sigma}_{{\mu}A}^h$ is the cross-section of $h$-shower generation, ${\nu}_n^h$ is neutron multiplicity in a $h$-shower. The cross-section can be written as 
\begin{equation}
{\sigma}_{{\mu}A}^h = {\sigma}_{{\mu}N}A^{\rho},
\label{eq7}
\end{equation}
here ${\sigma}_{{\mu}N}$  is a cross-section of deep-inelastic ${\mu}N$-interaction, ${\rho}$ is a degree of nucleon shadowing in a nucleus for virtual photons.

In the energy $E_{\mu}$ range of $10 - 10^4$ GeV the cross-section ${\sigma}_{{\mu}N}$ is constant: ${\sigma}_{{\mu}N}=1.1{\times}10^{-28}$ cm$^2$. 
In accordance with the experimental data deep-inelastic photonuclear interaction of cosmic muons is characterized by value 
${\rho}$ = 0.96 \cite{May75}, \cite{Nik77}.
One can transform expression (\ref{eq6}), using formula (\ref{eq7}) and setting ${\rho} = 1$:
\begin{eqnarray}
Y_n^h(E_{\mu},A)=\frac{N_0}{A}{\langle}{\sigma}_{{\mu}N}A^{\rho}{\nu}_n^h{\rangle}{\approx} \nonumber\\
{\approx}N_0{\sigma}_{{\mu}N}{\langle}{\nu}_n^h{\rangle}=N_{{\mu}N}{\langle}{\nu}_n^h{\rangle} ~{\mathrm{(cm^2/g)}}.
\label{eq8}
\end{eqnarray}
The number of ${\mu}N$-interactions $N_{{\mu}N}$ does not depend on $E_{\mu}$ and, practically, on $A$. Hence, the dependence of the 
yield $Y_n^h$ on $E_{\mu}$  and $A$ is contained in the ${\langle}{\nu}_n^h{\rangle}$ value. As follows from the experiments 
\cite{Pau62}, \cite{Bar72}, \cite{Gri58}, in deep-inelastic collisions of a particle with a nucleus the neutron number $n_n$ weakly correlates with the 
particle energy and mostly depends on $A$.
Therefore, the multiplicity ${\langle}{\nu}_n^h{\rangle}$ is defined by a multiplicity of pions ${\nu}_{\pi}^\pm$ and the neutron number $n_n$:
\begin{equation}
{\langle}{\nu}_n^h(E_{\mu},A){\rangle}={\langle}{\nu}_{\pi}^{\pm}(E_{\mu},A){\rangle}n_n(A).
\label{eq9}
\end{equation}
The ${\langle}{\nu}_{\pi}^{\pm}{\rangle}$ value determines the yield $Y_{\pi}^{\pm}$ of charged pions in $h$-shower:
\begin{equation}
Y_{\pi}^{\pm}=\frac{N_0}{A}{\langle}{\sigma}_{{\mu}N}A^{\rho}{\nu}_{\pi}^{\pm}{\rangle}{\approx}N_{{\mu}N}{\nu}_{\pi}^{\pm} ~{\mathrm{(cm^2/g)}}.
\label{eq10}
\end{equation}
The dependence of the multiplicity ${\nu}_{\pi}^{\pm}$ on $E_{\mu}$ and $A$ can be factorized:
\begin{equation}
{\nu}_{\pi}^{\pm}(E_{\mu},A)=c_{\pi}E_{\mu}^{{\alpha}_{\pi}}A^{{\beta}_{\pi}}.
\label{eq11}\\
\end{equation}
Substituting (\ref{eq11}) in (\ref{eq10}), we obtain an expression for the pion yield:
\begin{equation}
Y_{\pi}^{\pm}(E_{\mu},A)= N_{{\mu}N}c_{\pi}E_{\mu}^{{\alpha}_{\pi}}A^{{\beta}_{\pi}}.
\label{eq12}\\
\end{equation}
Taking into account (\ref{eq8}),(\ref{eq9}) and (\ref{eq12}) we get:
\begin{equation}
Y_n^h=Y_{\pi}^{\pm}(E_{\mu},A)n_n(A).
\label{13}
\end{equation}

\section {IV. Correlation between the yield $Y_{\pi}^{\pm}$ and the muon nuclear energy loss. The phenomenological expression for the yield $Y_{\pi}^{\pm}$}

Energy to produce pions is a portion  of the muon nuclear energy loss $b_n$, and energy for neutron production is taken from the 
shower charged pions. Hence, the yields $Y_{\pi}^{\pm}$, $Y_n^h$ are associated with the $b_n$ loss:
\begin{equation}
b_n=\frac{N_0}{E_{{\mu}A}}\int_0^{E_{\mu}}{\varepsilon}_hd{\sigma}(E_\mu,{\varepsilon}_h).
\label{14}
\end{equation}
Here ${\varepsilon}_h$ is  energy transferred  by muon to  $h$-shower, that is loss $b_n$  are connected with generation of $h$-showers only. 
This is valid for ultrarelativistic muons. Passing in (\ref{14}) to the mean muon energy transfer ${\overline {\varepsilon}}_h$  and using formula (\ref{eq7}), at ${\rho}=1$ we get:
\begin{equation}
b_n=N_0\frac{{\overline {\varepsilon}}_h}{E_\mu}{\sigma}_{{\mu}N}=N_{{\mu}N}\frac{{\overline {\varepsilon}}_h}{E_\mu} {\mathrm{(cm^2/g)}}.
\label{15}
\end{equation}
The value $b_n = 4.0{\times}10^{-7}$ cm$^2$/g  is constant in the $E_{\mu}$  range from 10 to $10^4$ GeV.  
Consequently, the ratio ${\overline {\varepsilon}_h/E_{\mu}}$  is constant too.

Expressions (\ref{15}) and (\ref{eq10}) have the same shape and dimension with the difference that the multiplicity 
${\langle}{\nu}_{\pi}^{\pm}{\rangle}$ is the number of charged pions in the shower  ${\overline {\varepsilon}}_h$ only. The number of charged pions
is connected with $E_{\mu}$ by dependence $E_{\mu}^{{\alpha}_{\pi}}$  \cite{Bar52}, \cite{Gru72}.
Multiplying both sides of equation (\ref{15}) by $E_{\mu}^{{\alpha}_{\pi}}$, we obtain the energy of the charged component of the muon nuclear energy loss
\begin{equation}
b_nE_{\mu}^{{\alpha}_{\pi}} = N_{{\mu}N}(\frac{{\overline {\varepsilon}}_h}{E_{\mu}}E_{\mu}^{{\alpha}_{\pi}}),
\label{17}
\end{equation}
in which the value of $\frac{{\overline {\varepsilon}}_h}{E_{\mu}}E_{\mu}^{{\alpha}_{\pi}}$ gives the energy of the charged component of the shower 
${\overline {\varepsilon}}_h$ contained  ${\langle}{\nu}_{\pi}^{\pm}{\rangle}$ pions.
This energy is distributed among ${\pi}^{\pm}$-pions in acts of deep-inelastic ${\pi}N$-scattering. Neglecting ${\pi}^{\pm}$-decays in flight, 
we can assume that the charged component of $h$-shower loses all its energy through ionization $({\varepsilon}_{\pi}^{ion})$, disintegration of 
nuclei in ${\pi}A$-interactions $(E_{dis})$ and generation of charged pion mass $(m_{\pi}c^2)$.

The value ${\varepsilon}_{1\pi}^{ion}$ is the pion energy loss over the mean free path ${\lambda}_\pi$ for inelastic $\pi A$-reactions. The length ${\lambda}_\pi$ is not 
connected practically with energy of a pion and weakly depends on $A$. The $E_{dis}$ magnitude varies in a similar way \cite{Pau62}, \cite{Gri58}.
Energy expended per a pion can be expressed as the sum ${\varepsilon}_{1\pi}^{ion}+E_{dis}+m_{\pi}c^2 = {\varepsilon}_{1\pi}(A)$, then
\begin{equation}
{\varepsilon}_{1\pi}(A){\langle}{\nu}_{\pi}^{\pm}(E_{\mu},A){\rangle} = \frac{{\overline {\varepsilon}}_h}{E_{\mu}}E_{\mu}^{{\alpha}_{\pi}}.
\label{18}
\end{equation}
Using the expression (\ref{eq11}) for ${\nu}_{\pi}^{\pm}$, we obtain the equality
\begin{equation}
c_{\pi}E_{\mu}^{{\alpha}_{\pi}}A^{{\beta}_{\pi}}=\frac{{\overline {\varepsilon}}_h}{E_{\mu}}E_{\mu}^{{\alpha}_{\pi}}\frac{1}{{\varepsilon}_{1\pi}(A)},
\label{19}
\end{equation}
whence it follows that $c_{\pi}=\frac{{\overline {\varepsilon}}_h}{E_{\mu}}$ and ${\varepsilon}_{1\pi}$ depends on $A$ in the following way:
\begin{equation}
{\varepsilon}_{1\pi} = 1/A^{{\beta}_\pi} ~{\mathrm{(GeV)}}.
\label{20}
\end{equation}

The multiplicity ${\langle}\nu_\pi^\pm{\rangle}$ weakly depends on $A$ and type of a particle - projectile. The dependence ${\langle}\nu_\pi^{\pm}(A){\rangle}$ in the 
form $A^{{\beta}_\pi}$ at $\beta_{\pi}= 0.14 {\pm} 0.03$ was obtained in the experiment described in Ref. \cite{Mey63} for protons with an 
energy of $20 - 27$ GeV; the value ${\beta}_{\pi} = 0.13 {\pm} 0.02$ was defined for ${\pi}^-$-mesons at energy of 17 GeV 
in Ref.\cite{Koh67}.
Dependence $\nu_{\pi}^\pm$ on $A$ is caused due to pion multiplication within a nucleus which  results in a decrease in the 
${\overline \varepsilon}_{1\pi}^{ion}$ and ${\overline E}_{dis}$ values and an increase in the fraction of the shower energy going to the pion production. The role of this process increases at increasing of $A$, that leads to an inverse $A$ dependence of ${\varepsilon}_{1\pi}$.
Taking into account (\ref{20}), we obtain for ${\langle}\nu_\pi^\pm{\rangle}$:
\begin{equation}
{\langle}\nu_\pi^\pm (E_\mu,A){\rangle}= \frac{{\overline {\varepsilon}}_h}{E_{\mu}}E_{\mu}^{{\alpha}_{\pi}}A^{{\beta}_\pi}.
\label{21}
\end{equation}
Substituting (\ref{21}) into (\ref{eq10}) and using (\ref{15}), we arrive at the expression
\begin{equation}
Y_\pi^\pm (E_\mu,A)=b_nE_{\mu}^{{\alpha}_\pi}A^{\beta_\pi}.
\label{22}
\end{equation}
The value of the exponent ${\alpha}_\pi=0.75$ was defined for the first time in the EAS \cite{Bar52} and then confirmed by 
calculations \cite{Gru72}. Assuming ${\alpha}_\pi = 0.75$ and ${\beta}_\pi = 0.13$, we obtain the expression for the yield $Y_\pi^\pm$:
\begin{equation}
Y_\pi^\pm = b_n E_\mu^{0.75}A^{0.13}.
\label{23}
\end{equation}

In Ref. \cite{Wan01}, the yield $Y_\pi^+$ value for liquid scintillator (LS) was obtained using the Monte Carlo package FLUKA:
 \begin{equation}
Y_\pi^+ = 4.45{\times}10^{-7} E_\mu ^{0.80}.
\label{25}
\end{equation}
In Ref. \cite{Del95} the yield $Y_\pi^+$  
for LS has been calculated analytically at the depths of 20, 100, 500 m w.e. to which  energies ${\overline E}_\mu$ of 10.3, 22.4, and 80 GeV 
were attributed in Ref. \cite{Wan01}. One can define the values of the yield $Y_\pi^+$ in LS (${\overline A}=10.3$), using different 
formulae at $E_\mu$ = 80 GeV: $Y_\pi^+ = 0.86{\times}10^{-5}$ \cite{Del95};   $1.48{\times}10^{-5}$  \cite{Wan01};  
$Y_\pi^+ = \frac{1}{2} Y_\pi^{\pm} = 0.72{\times}10^{-5}$ (using formula (\ref{23}) while assuming $Y_\pi^+= Y_\pi^-$). 
The scatter of the values obtained demonstrates significant uncertainties given calculations. One can add that the $\alpha$ value obtained by various 
authors using the Monte Carlo method is within a range from 0.6 to 0.8.

\section {V. The phenomenological expression for the yield $Y_n^h$}

The dependence of the yield $Y_n^h$ on $E_\mu$ and $A$ is contained 
in the ${\langle}\nu_n^h{\rangle}$ value which can be factorized:
 \begin{equation}
{\langle}\nu_n^h(E_\mu,A){\rangle} = c_nE_\mu^{{\alpha}_n}A^{\beta}. 
\label{26}\\
\end{equation}
According to (\ref{eq9}) and (\ref{21}), the multiplicity ${\langle}\nu_n^h{\rangle}$  can be represented as
 \begin{equation}
{\langle}\nu_n^h(E_\mu,A){\rangle} = {\langle}\nu_\pi^\pm{\rangle}n_n(A)=\frac{{\overline {\varepsilon}}_h}{E_{\mu}}E_{\mu}^{{\alpha}_{\pi}}A^{{\beta}_\pi}n_n(A).
\label{27}\\
\end{equation}
The right-hand sides of the equations (\ref{26}) and (\ref{27}) are equal to each other:
$c_nE_\mu^{{\alpha}_n}A^{\beta} =\frac{{\overline {\varepsilon}}_h}{E_{\mu}}E_{\mu}^{{\alpha}_{\pi}}A^{{\beta}_\pi}n_n(A)$.
Since the $n_n$ value is not dependent on energy $E_\mu$, then ${\alpha}_n={\alpha}_\pi$ and 
$c_nA^{\beta - \beta_\pi} = \frac{\overline {\varepsilon}_h}{E_\mu}n_n(A)$. Hence it follows: 
$c_n = \frac{\overline {\varepsilon}_h}{E_\mu}$  and $n_n(A) = A^{\beta - \beta_\pi}$. 
Denoting $\beta - \beta_\pi = \beta_n$, substituting $n(A) =A^{\beta_n}$ in the expression for ${\nu}_n^h$, 
and taking into account (\ref{13}) and (\ref{22}), we obtain $Y_n^h(E_\mu,A) = b_nE_\mu^{\alpha_\pi}A^{\beta_\pi}A^{\beta_n}$.

Experimental data and calculations within the INC model \cite{Bar72} are in good agreement with exponent $\beta_n = 0.74 \pm 0.10$. 
Using the value of $\beta_\pi = 0.13$ and taking into account the uncertainties of definition of the $\beta$ values 
one can assume $\beta_\pi  + \beta_n  \approx 0.90$. In such a case we get the expression:
\begin{equation}
Y_n^h(E_\mu,A)=b_nE_\mu^{0.75}A^{0.90}.
\label{28}\\
\end{equation}

\section {VI. The phenomenological expression for the yield $Y_n^{\mathrm{em}}$}

Muon initiates an em-shower via $\delta$-electron, radiative $\gamma$-quantum (r) or $e^+e^-$-pair (p). 
The em-shower produces a low neutron amount, but due to a high generation cross-section 
the em-showers provide contribution to the cg-neutron yield comparable with that from $h$-showers. Any em-shower consists of electrons $e^+, e^-$ 
and shower $\gamma$-quanta (photons). 
Amounts of both shower charged particles $N_{sh}^e$ and photons $N_{sh}^\gamma$ are proportional to the shower energy $E_{\mathrm{em}}$. 
The number of photons with an energy above 10 MeV is 2 - 3 times the number $N_{sh}^e$.
At high energies $E_{\mathrm{em}}$, hadron $h$-subshowers appear in the em-shower structure, which are produced via photoproduction. The probability of this process 
is low due to the steep shower photon spectrum $P(\varepsilon_\gamma) \propto 1/{\varepsilon}_{\gamma}^2$. Contribution of $h$-subshowers to the value of the yield 
$Y_n^{\mathrm{em}}$ will not be considered below.
In contrast to the $h$-showers practically all the em-shower energy is spent for a medium ionization. 

The dominant neutron production process in em-showers 
is photoproduction because, firstly, the photoproduction cross section is $\sim 10^2$ times the cross-section of the $eA$-electronuclear reactions 
and, secondly, $N_{sh}^\gamma {>} N_{sh}^e$.
Among photoproduction processes, the largest contribution to the yield $Y_n^{\mathrm{em}}$ is introduced by GDR producing $ev$-neutrons.  
The GDR region is within the range from the nucleon binding energy in the nucleus up to $\sim 20$ MeV. The GDR maximum is given 
by expression  $E_{\gamma}^{max} \approx 40 A^{-0.2}$ MeV. Photoabsorption cross-section $\sigma_a$  is given by: 
\begin{equation}
\sigma_a  = \int_0^{m_{\pi}c^2} \sigma_{\gamma A}dE_{\gamma} \approx 60 \frac{(A-Z)Z}{A}10^{-27}~ {\mathrm{cm^2 MeV}}.
\label{29}\\
\end{equation}
Due to the large GDR width (2 to 8 MeV) and its maximum location the photoneutron yield weakly depends on the shape of the photon 
spectrum $P(\varepsilon_\gamma)$  and it is determined by the number of photons: 
$Y_n^{\mathrm{em}} \propto N_{sh}^\gamma \propto E_{\mathrm{em}}$. Since $N_{sh}^\gamma \propto E_{\mathrm{em}}$, and the em-shower generation is determined 
by the cross-section $\sigma_{\mu A}^{\mathrm{em}}$, the $Y_n^{\mathrm{em}}$ yield is proportional to the em-muon energy loss:
\begin{equation}
Y_n^{\mathrm{em}} \propto (\frac{dE_\mu}{dx})^{\mathrm{em}} = k_\delta + (b_r(A)+b_p(A))E_\mu,
\label{30}\\
\end{equation}
here $k_\delta, b_r, b_p$ are functions weakly dependent on the $E_\mu$. The $k_\delta$ value at $E_\mu$ above 10 GeV increases insignificantly 
and is practically independent on $A$. So one can assume that $k_\delta \approx const$. Values $b_r$ and $b_p$ represent the muon energy loss:
\begin{equation}
b_{r,p}=( \frac{1}{E_\mu}\frac{dE_\mu}{dx} )_{r,p} = \frac{N_0}{E_\mu A} \int_0^{E_\mu} \varepsilon_{r,p} \sigma_{r,p}(E_\mu, \varepsilon)d\varepsilon,
\label{32}\\
\end{equation}
here $\varepsilon_{r,p}$ is $\gamma$-quantum or pair energy, $\sigma_{r,p}$ is cross-section of respective process.\\
The loss $b_r$ and $b_p$ within the range from $E_\mu$ 40 to 400 GeV are practically independent on energy $E_\mu$,  
in this case $( \frac{dE_\mu}{dx})_{r,p} \propto E_\mu$. These loss are connected with the matter properties by the following dependence:
\begin{equation}
b_{r,p}(A) \propto Z^2 /A \approx A^{0.94}/4 \propto A^{0.94} \approx A^{1.0}.
\label{33}\\
\end{equation} 
Having introduced into (\ref{30}) the coefficient $\nu_n^{\gamma A} (A)$, which considers a neutron multiplicity at the $\gamma A$-absorption, and
also $E_\mu$ dependence, we obtain the expression
\begin{equation}
Y_n^{\mathrm{em}}=c^{\mathrm{em}} \nu_n^{\gamma A}(k_\delta +b_r (A)E_\mu^{1.0} + b_p(A)E_\mu^{1.0}),
\label{34}
\end{equation}
where $c^{\mathrm{em}}$ is a portion of em-loss for producing neutrons, which is the same for all em-processes.

The neutron production function was approximated in the GDR region by expression 
$\sigma_a \nu_n^{\gamma A}$ = $5.2 \times 10^{-4} A^{1.8}$ MeV{\em barns} \cite{Jon53}. 
Comparing this formula with (\ref{29}) and assuming $(A-Z)Z/A \approx A^{1.0}/4 \propto A^{1.0}$, we arrive at dependence $\nu_n^{\gamma A} = c_{\gamma A}A^{0.8}$
which characterizes a photoneutron multiplicity in em-showers at any energy $E_{\mathrm{em}}$.

One can transform the expression (\ref{34}), in accordance with (\ref{33}) assuming $b_r(A) \approx a_rA^{1.0}$, $b_p(A) \approx a_pA^{1.0}$ (the 
values $a_r$ and $a_p$ are constants) and using the expression $\nu_n^{\gamma A}=c_{\gamma A}A^{0.8}$:
\begin{eqnarray}
Y_n^{\mathrm{em}}(E_\mu,A)=
c^{\mathrm{em}}c_{\gamma A}k_\delta A^{0.8}+\nonumber\\
c^{em}c_{\gamma A}a_rA^{1.8}E_\mu^{1.0}+c^{\mathrm{em}}c_{\gamma A}a_pA^{1.8}E_\mu^{1.0}.
\label{35}
\end{eqnarray}
Joining the constants in (\ref{35}) in the $c_\delta, c_r, c_p$ coefficients
we obtain the dependence of the $Y_n^{\mathrm{em}}$ yield on $E_\mu$ and $A$:
\begin{equation}
Y_n^{\mathrm{em}}(E_\mu,A)=c_\delta A^{0.8}+c_rA^{1.8}E_\mu^{1.0}+c_pA^{1.8}E_\mu^{1.0}.
\label{36}
\end{equation}
In this expression representing the neutron yield for em-processes only one can include the $Y_n^v$  term relating to the nuclear muon loss
and corresponding to neutron production by virtual photons.
In spite of the more rigid spectrum $\propto 1/E_\gamma^v$ in contrast to spectrum of real photons in the em-showers, virtual photons produce the 
overwhelming majority of neutrons also via GDR due to its large width. As a result, the expression for the $Y_n^v$ takes a form similar to 
the expression for the neutron yield in $\delta$-showers: $Y_n^v=c_vA^{0.8}$. Including this formula to (\ref{36}) 
we obtain the neutron yield in all the processes except for $h$-showers:
\begin{equation}
Y_n^{ph}=(c_\delta + c_v)A^{0.8}+(c_r+c_p)A^{1.8}E_\mu^{1.0}
\label{38}
\end{equation}
Members of this expression represent the neutrons produced via nuclear photoeffect.
These neutrons originate from primary nuclei $A$ of the matter in contrast to the $h$-showers, where evaporative neutrons are emitted by remnants of the nuclei $A_r$
Starting from energy of $E_\mu \sim 100$ GeV, the second term dominates in the yield (\ref{38}), so the $Y_n^{ph}$ yield can be represented 
in a form similar to expression $Y_n^h$ (\ref{28}): $Y_n^{ph} = cE_\mu^\alpha A^{\beta_{ph}}$. Here exponents $\alpha$ and $\beta_{ph}$ are slightly 
less than 1.0 and 1.8, respectively. Thus, the total neutron yield is a sum of  components $Y_n^h$ and $Y_n^{ph}$:
\begin{equation}
Y_n \approx Y_n^h + Y_n^{ph} = b_nE_\mu^{0.75}A^{0.90}+cE_\mu^\alpha A^{\beta_{ph}}.
\label{40}
\end{equation}
Substituting $Y_n^{\mathrm{UF}}$ for $Y_n$ in (\ref{40}), we obtain:
\begin{equation}
b_n^{\mathrm{UF}}E_\mu^{0.78}A^{0.95}=b_nE_\mu^{0.75}A^{0.90}+cE_\mu^\alpha A^{\beta_{ph}}.
\label{41}
\end{equation}

Using the expressions $Y_n^h$  and $Y_n^{\mathrm{UF}}$, one can define the portion of the hadron component in the total yield $Y_n$ as follows:
$K(E_\mu,A)$ =$Y_n^h/Y_n^{\mathrm{UF}}=0.91(E_\mu)^{-0.03}A^{-0.05}$.
For example, at $E_\mu = 280$ GeV the $K$ values are enclosed within 0.68 and 0.59 for  numbers $A$ from 12 to 207.

UF parameters were fitted to experimental data. This procedure takes into account contribution of the $Y_n^{\mathrm{em}}$ component 
into a total cg-neutron yield as well as an impact of the real muon spectrum  on the real  $Y_n$ value. 
The $\sigma_{{\mu}A}\nu_n$ function in equation (\ref{eq1}) is not only summary for the $\mu A$-interactions but also integrated over the muon 
spectrum at a depth of observation.
Due to the cg-neutron yield energy dependence $E_\mu^{\alpha}$ and a quasiflat muon spectrum deep underground 
$\frac{dN_\mu}{dE} \sim \frac{1}{(E_0(H)+E_\mu)^\gamma}$, the use of monoenergetic muons with energy ${\overline E}_\mu$  
in calculations results in the $Y_n$ yield value overestimated by 12{\%} for ${\overline E}_\mu \sim 100$ GeV and 5{\%} 
for ${\overline E}_\mu \sim 300$ GeV if $\alpha =0.75$ \cite{Hag00}, \cite{Hei02}.
Nevertheless, the measured yield $Y_n$ is attributed to energy ${\overline E}_\mu$ since the ${\overline E}_\mu$ value is a 
natural physical parameter characterizing the muon flux and muon interactions underground.

It can be noted that in the high energy $h$-shower a large number of neutrons is produced. This is a rare event leading
 to significant fluctuations in the value of $Y_n$ obtained during a finite-time measurement. Thus, 
the yield calculated by the UF is an asymptotic value for the yield magnitudes which are obtained in measurements.

\section {VII. Conclusion}

Empirical expressions $c_A E_\mu^\alpha$ and $c_E A^\beta$  are the simplest 
representations of the cg-neutron yield dependence on  $E_\mu$ and $A$. Obtained by fitting to the experimental or calculated data, 
they reflect trends in the values $Y_n(E_\mu)$ and $Y_n(A)$ without discovering their correlation with physical processes of the neutron production by muons. 
Universal formula $Y_n^{\mathrm{UF}}=b_n^{\mathrm{UF}}E_\mu^\alpha A^\beta$  is also empirical due to the method of its derivation. But UF uncovers 
the meaning of the coefficients $c_A, c_E$ and points out that the neutron yield is connected with muon energy loss.
The UF kernel is the phenomenological $Y_n^h$ expression which is obtained within the framework of
 the concept of deep-inelastic muon scattering and $\pi A$-interaction.
This approach allows to associate the yield $Y_n^h$ with the muon nuclear energy loss and the characteristics of neutron 
production in the hadron showers and to explain the origin of the exponent values $\alpha, \beta$  in the $Y_n^h$ and $Y_n^{\mathrm{UF}}$ expressions.

\end{document}